\documentclass[runningheads]{llncs}
\usepackage[utf8]{inputenc}

\usepackage[T1]{fontenc}
\usepackage{newtxtext,newtxmath}
\usepackage[scaled=0.80]{beramono}
\usepackage{microtype}
\usepackage{csquotes}

\usepackage{upquote}
\newcommand\fullyjustified{\parfillskip 0pt}

\usepackage{listings}
\newcommand\codefont{\ttfamily\fontsize{7.5}{8.5}\selectfont}
\lstdefinelanguage{SPARQL}{
  keywords={SELECT, WHERE, OPTIONAL},
  comment=[f]{\#},
}
\lstdefinelanguage{Turtle}{
  comment=[f]{\#},
}
\usepackage{array}
\let\currentfirstcolstyle\relax
\newcommand\firstcolstyle[1]{%
  \gdef\currentfirstcolstyle{#1}%
  #1\ignorespaces%
}
\newcommand\rowstyle[1]{%
  \gdef\currentrowstyle{#1}%
  #1\ignorespaces%
}
\newcolumntype{+}{>{%
  \global\let\currentrowstyle\relax%
  \currentfirstcolstyle%
}}
\newcolumntype{^}{>{%
  \currentrowstyle%
}}

\usepackage[
  font=footnotesize,
  labelfont=bf,
  singlelinecheck=false,
]{caption}
\usepackage[sort]{cite}
\usepackage[hyphens]{url}

\usepackage[hidelinks]{hyperref}
\usepackage[capitalize,noabbrev,nameinlink]{cleveref}

\usepackage{xspace}
\newcommand\Acronym[1]{%
  \expandafter\def\csname#1\endcsname{{\scshape #1}\xspace}%
  \expandafter\def\csname#1s\endcsname{{\scshape #1}s\xspace}%
}
\Acronym{api}
\Acronym{html}
\Acronym{http}
\Acronym{ltbqp}
\Acronym{rdf}
\Acronym{shacl}
\Acronym{sparql}
\newcommand\dbpedia{\textsc{db}pedia\xspace}

\newcommand\LinkingStructures{\ensuremath{\mathcal{LS}}\xspace}
\newcommand\Reachable{\ensuremath{D}\xspace}
\newcommand\RelevanceCriteria{\ensuremath{\mathcal{RC}}\xspace}
\newcommand\Seeds{\ensuremath{\mathcal{S}}\xspace}
\newcommand\Triples{\ensuremath{\mathcal{T}}\xspace}
\newcommand\Web{\ensuremath{W}\xspace}
\newcommand\doc{\ensuremath{d}\xspace}
\newcommand\evaluation[2]{\ensuremath{[\![#1]\!]_{#2}}}

\newcommand\getLinkingStructure{\ensuremath{\mathit{obtainLS}}\xspace}
\newcommand\guided{\ensuremath{\mathit{g}}\xspace}
\newcommand\linkingstructure{\ensuremath{\mathit{ls}}\xspace}
\newcommand\parse{\ensuremath{\mathit{parse}}\xspace}
\newcommand\query{\ensuremath{Q}\xspace}
\newcommand\reachability{\ensuremath{\mathit{r}}\xspace}

\newcommand\reachable{\ensuremath{\mathit{reachable}}\xspace}
\newcommand\relevancecriterion{\ensuremath{\mathit{rc}}\xspace}
\newcommand\triplepattern{\ensuremath{\mathit{tp}}\xspace}
\newcommand\triple{\ensuremath{t}\xspace}
\newcommand\true{\ensuremath{\mathit{true}}\xspace}

\usepackage[dvipsnames,svgnames]{xcolor} % text color
\usepackage[normalem]{ulem} % wavy underlines

\makeatletter
\font\uwavefont=lasyb10 scaled 700
\def\spelling{\bgroup\markoverwith{\lower3.5\p@\hbox{\uwavefont\textcolor{Red}{\char58}}}\ULon}
\def\grammar{\bgroup\markoverwith{\lower3.5\p@\hbox{\uwavefont\textcolor{LimeGreen}{\char58}}}\ULon}
\def\phrasing{\bgroup\markoverwith{\lower3.5\p@\hbox{\uwavefont\textcolor{RoyalBlue}{\char58}}}\ULon}

\newcommand\remove{\bgroup\markoverwith{\textcolor{red}{\rule[0.5ex]{2pt}{0.4pt}}}\ULon}
\makeatother

\begin{document}

\title{%
  Guided Link-Traversal-Based Query Processing
  \vspace{-.5\baselineskip}
}
\author{%
  Ruben Verborgh%
  \and
  Ruben Taelman%
  \vspace{-.75\baselineskip}
}
\institute{%
  IDLab,
  Dept.\ of Electronics and Information Systems,
	UGent -- imec,
  Belgium
}
\maketitle

\vspace{-1.5\baselineskip}

\begin{abstract}
{
\fullyjustified
% Context:      Why the need is so pressing or important
Link-Traversal-Based Query Processing (\ltbqp)
is a~technique for evaluating queries over a~web of data
by starting with a~set of seed~documents
that is dynamically expanded through following hyperlinks.
% Need:         Why something needed to be done at all
Compared to query evaluation over a~static set of sources,
\ltbqp is significantly slower because of the number of needed network requests.
Furthermore, there are concerns regarding relevance and trustworthiness of results,
given that sources are selected dynamically.
% Task:         What was undertaken to address the need
To~address both issues,
we propose \emph{guided} \ltbqp,
a~technique in which information about
document linking structure and content policies
is passed to a~query processor.
Thereby,
the processor can prune the search~tree of documents
by only following relevant links,
and restrict the result set to desired results
by limiting which documents are considered for what kinds of~content.
% Object:       What the present document does or covers
In this exploratory paper,
we describe the technique at a~high level
and sketch some of its applications.
% Findings:     What the work done yielded or revealed
We argue that such guidance
% Conclusion:   What the findings mean for the audience
can make \ltbqp a~valuable query strategy
in decentralized environments,
where data is spread across documents
with varying levels of user~trust.
% Perspectives: What the future holds, beyond this work

}
\end{abstract}

\section{Querying Data on the Web}
\label{sec:Introduction}
{
\fullyjustified
Link-Traversal-Based Query Processing (\ltbqp)
enables the evaluation of \sparql queries
over hyperlinked \rdf documents on the Web
rather than \rdf~databases.
Evaluating \sparql over the Web of Data
is supported by a~family of \emph{reachability semantics}~\cite{Hartig2013},
where the source of the \rdf data used for a~query
is an expansible set of \emph{seed documents}
to which new documents are added dynamically
based on the already obtained \rdf~triples.

\ltbqp suffers from performance problems
compared to \sparql query evaluation over one or more \rdf~databases,
given the dynamic nature of its sources
and the impossibility of predicting
which reachable documents
will contribute to the result~set.
While heuristics can \emph{prioritize} certain links
in order to reduce response times~\cite{WalkingWithoutMap},
they cannot safely \emph{prune} the document search tree.
Given the Web's openness,
issues such as trustworthiness and license compliance emerge
when incorporating unknown documents.

These problems can be addressed
by tailoring queries to a~specific user and context.
For~instance, the query language could explicitly express
which traversals are permitted~\cite{LDQL}.
However, we aim to maintain independence
between applications on the one~hand,
and the structure of the data~network
and the user's content preferences on the other~hand.
Therefore,
we introduce the concept of
\emph{Guided} Link-Traversal-Based Query Processing (Guided \ltbqp),
in which a~user-controlled context guides the query engine's process.

}

\section{Example Use Case}
\label{sec:UseCase}

\begin{figure}[tb]
\begin{minipage}[t]{.5\linewidth}
  \begin{lstlisting}[
  label={qry:Friends},
  caption={%
    Application query in \sparql
  },
  language=SPARQL,
]
SELECT ?friend ?name ?email ?picture WHERE {
  <https://uma.ex/#me> foaf:knows ?friend.
  ?friend foaf:name ?name.
  OPTIONAL { ?friend foaf:mbox ?email.
             ?friend foaf:img  ?picture. }
}
\end{lstlisting}

\begin{lstlisting}[
  label={lst:Uma},
  caption={%
    \rdf from \url{https://uma.ex/}
  },
]
<https://uma.ex/#me> foaf:knows
  <https://ann.ex/#me>, <https://bob.ex/#me>.
<https://bob.ex/#me> foaf:img <bob.jpg>.
\end{lstlisting}

\begin{lstlisting}[
  label={lst:Ann},
  caption={%
    \rdf from \url{https://ann.ex/}
  },
]
<https://ann.ex/#me> foaf:isPrimaryTopicOf <https://ann.ex/about/>.
<https://ann.ex/#me> foaf:weblog <https://ann.ex/blog/>.
<https://ann.ex/#me> foaf:maker <https://photos.ex/ann/>.
\end{lstlisting}
\end{minipage}
\begin{minipage}[t]{.5\linewidth}
\begin{lstlisting}[
  label={lst:Bob},
  caption={%
  \rdf from \url{https://bob.ex/}
  },
]
<https://bob.ex/#me> foaf:name "Bob";
  foaf:email <mailto:me@bob.ex>;
  foaf:img <funny-fish.jpg>.
<https://uma.ex/#me> foaf:knows
  <http://dbpedia.org/resource/Mickey_Mouse>.
<https://ann.ex/#me> foaf:name "Felix";
\end{lstlisting}

\begin{lstlisting}[
  label={lst:AnnDetails},
  caption={%
    \rdf from \url{https://ann.ex/about/}
  },
]
<https://ann.ex/#me> foaf:name "Ann";
  foaf:email <mailto:me@ann.ex>;
  foaf:img <ann.jpg>.
\end{lstlisting}
\end{minipage}
\end{figure}

{
\fullyjustified
We will discuss an example using personal Linked Data~\cite{verborgh_timbl_chapter_2020},
for which document-based data organization is common.
Consider an address book app displaying contacts,
operated by a~user Uma with identifier \url{https://uma.ex/#me}.
Its data demand is captured by the query in \cref{qry:Friends}.
The contents of Uma's own profile are displayed in \cref{lst:Uma}.
Note how Uma's contact Ann (\cref{lst:Ann})
maintains her personal details in a~separate document (\cref{lst:AnnDetails}).
Contact Bob (\cref{lst:Bob}) is a~self-professed jokester
who picks a~funny picture for himself,
declares Mickey Mouse to be Uma's friend,
and Ann's name to be \enquote{Felix}.

}

With traditional link-traversal-based query evaluation
under several (but not all) reachability semantics~\cite{Hartig2012},
a~query engine could fetch at least 7~documents:
profile documents of 3~people (Uma, Ann, Bob),
3~documents referred to by Ann's profile (\cref{lst:AnnDetails}),
and the \dbpedia entry on Mickey Mouse.
Results could include those in~\cref{tbl:Results}.

\begin{table}[tb]
  \tabcolsep 3pt
  \codefont
  \begin{tabular}{|+r|^l|^l|^l|^l|}
    \hline
    \rowstyle{\bfseries}
    \firstcolstyle{\bf}
      & ?friend               & ?name             & ?email             & ?picture                        \\
    \hline
    1 & <https://ann.ex/\#me> & "Ann"             & <mailto:me@ann.ex> & <https://ann.ex/about/ann.jpg>  \\
    2 & <https://bob.ex/\#me> & "Bob"             & <mailto:me@bob.ex> & <https://uma.ex/bob.jpg>        \\
    3 & <https://bob.ex/\#me> & "Bob"             & <mailto:me@bob.ex> & <https://bob.ex/funny-fish.jpg> \\
    4 & <https://ann.ex/\#me> & "Felix"           & <mailto:me@ann.ex> & <https://ann.ex/about/ann.jpg>  \\
    5 & dbr:Mickey\_Mouse     & "Mickey Mouse"@en & \itshape NULL      & \itshape NULL                   \\
    \hline
  \end{tabular}
  \smallskip
  \caption{%
    Possible results of \ltbqp of the query in \cref{qry:Friends}
    with \url{https://uma.ex/} as seed
  }
  \label{tbl:Results}
\end{table}

\section{Leveraging Document Linking Structure}
\label{sec:LinkStructure}
{
\fullyjustified
A~first issue with the evaluation of the query in \cref{qry:Friends}
is that more documents than needed are downloaded
in order to find the results,
because the query engine cannot distinguish
between relevant and irrelevant documents.
For example,
Ann splits her profile into 3~separate documents (\cref{lst:Ann}),
only one of which contains her contact details (\cref{lst:AnnDetails}),
the others listing blog posts and photos.
This contrasts with Uma and Bob (\cref{lst:Uma,lst:Bob}),
who place their contact details directly in their profile document.

Suppose the query engine is supplied with knowledge
about the linking structure of documents,
for instance through a~shape description language such as \shacl.
The shape could detail that
Ann stores her contact details
in the document referred to by the
\verb!foaf:isPrimaryTopicOf!~relation.
Likewise, for Uma and Bob,
their shape could express that
all of their contact details reside inside of their main profile document.
While the availability of such structural knowledge seems far-fetched on the public Web,
in the context of personal data spaces
within the Solid ecosystem~\cite{verborgh_timbl_chapter_2020},
this structure needs to be explicit anyway
so applications know where to \emph{write} new~data.
As such, users do not need to provide this along with queries,
since every person or domain can publish their~own.

}

Based on this structure,
the engine can decide
to skip the links to
\url{https://ann.ex/blog/},
\url{https://photos.ex/ann/},
and
\url{http://dbpedia.org/resource/Mickey_Mouse}
to collect contact details for Ann and Bob.
After all,
their data spaces indicate conformance to certain shapes,
so for Ann it is sufficient to follow
the \verb!foaf:isPrimaryTopicOf! link
to \url{https://ann.ex/about/},
and for Bob to remain on \url{https://bob.ex/}.
This reduces the number of required network requests
for Ann from 4 to~2.

Note that the set of query results obtained after link pruning
could be different from \cref{tbl:Results}
if the structural descriptions are not truthful,
leading to the query engine to skip documents with matching triples.
However, the user remains in full control
of which structural descriptions are to be~taken into account
by the query engine.

\section{Incorporating Content Policies}
\label{sec:ContentPolicies}
A~second issue is that,
within the query graph formed by the union
of the \rdf graphs of each traversed document,
all triples are considered to be equally applicable.
This conflicts with the real world,
in which we rely on certain sources for certain kinds of facts,
but not for~others.
For example,
Bob's profile document contains false statements about Uma and Ann,
and an unhelpful statement about his own profile picture.

User Uma might want to capture the notion
that, for statements about her relation to~others,
she only considers her own profile document authoritative.
In~contrast, she allows people to make statements
about their own name and other contact details.
Yet if she specifies a~profile picture for a~person,
she wants that to take precedence over others.

If a~query engine takes these preferences into account,
then the evaluation of the query from \cref{qry:Friends}
would yield only results the user considers relevant,
namely rows 1~and~2 from \cref{tbl:Results}.
Rows 3--5 are undesired,
since they are based on triples
that fall outside of the user-specified content policy,
which indicates what triples are to be considered.
Because of this restriction,
the document at \url{http://dbpedia.org/resource/Mickey_Mouse}
does not need to be retrieved,
as the triple mentioning it also falls outside of the policy.
Combined with the reductions obtained in \cref{sec:LinkStructure},
the query engine thus only needs to download 4~documents instead of~7
to find all answers the user deems relevant,
namely the 3~profile documents (\cref{lst:Uma,lst:Ann,lst:Bob}),
and Ann's details (\cref{lst:AnnDetails}).

This concept of content policies,
which filter what triples of a~given document are to be considered,
has several purposes in addition to end-user trust.
For example,
a~user could demand to skip documents with a~closed license
in order to avoid copyright issues,
skip documents marked as suspicious by fact checkers
to combat fake news,
or selectively allow publication metadata from suspicious documents
for reference reasons,
but ignore their data~triples that are regarded as nonfactual.

\section{Query Semantics}
\label{sec:QuerySemantics}
As a~simplification,
and borrowing notation from existing literature~\cite{Hartig2012},
the evaluation of a~\sparql query~\query
over a~web of documents~\Web
under \emph{reachability-based semantics}
is equal to the evaluation of~\query
over the union of all \rdf triples extracted from documents
that are transitively reachable from a~set of seed documents~\Seeds,
namely
$
  \evaluation{\query}{\Web}^\reachability
  =
  \evaluation{\query}{\Web^\reachability}
$
with
$
  \Web^\reachability
  =
  \{
    \triple \in \Triples
    \mid
    \triple \in \parse(\doc)
      \land
      \doc \in \reachable(\Seeds)
  \}
$.

We incorporate the concept of \emph{linking structure}
through descriptions of specific linking structures
$\linkingstructure \in \LinkingStructures$,
where $\linkingstructure(\doc, \doc', \triplepattern) = \true$
iff, starting from a~document~\doc,
the document~$\doc'$ should be considered
for matches to the triple pattern~\triplepattern.
The linking structure associated with a~given document~\doc
is retrieved via a~function \getLinkingStructure.
% All documents that are reachable from a~set of seeds
% by following links conforming to a~specific linking structure
% are given by the function \reachableLS.

We incorporate the concept of a~\emph{content policy}
as a~set of relevance criteria
$\relevancecriterion \in \RelevanceCriteria$,
where
$\relevancecriterion(\triple, \doc) = \true$
iff a~triple~\triple from a~document~\doc
is considered relevant.

This leads to an initial query semantics for guided \ltbqp.
Instead of~\query,
the user passes an augmented query
$
  \query'
  =
  (
    \query,
    \getLinkingStructure,
    \RelevanceCriteria
  )
$
to the query engine.
We consider the evaluation of~$\query'$
under \emph{guided link-traversal-based semantics}
as the evaluation of~\query
over the set containing every triple that
a)~is obtained from any reachable document adhering to the linking structure
and
b)~is part of the relevant set of triples for that document,
so
$
  \evaluation{\query'}{\Web}^\guided
  =
  \evaluation{\query}{\Web^\guided}
$
with
$
  \Web^\guided
  =
  \{
    \triple \in \Triples
    \mid
    \triple \in \parse(\doc)
      \land
      \doc \in \Reachable
      \land
      \exists \relevancecriterion \in \RelevanceCriteria :
        \relevancecriterion(\triple, \doc)
  \}
$
and where
$
  \Reachable
  =
  \Seeds
  \cup
  \{
    \doc
    \mid
    \exists \doc' \in \Reachable :
      \doc'\! \neq \doc
      \land
      \linkingstructure = \getLinkingStructure(\doc')
      \land
      \exists \triplepattern \in \query :
        \linkingstructure(\doc, \doc', \triplepattern)
  \}
$.

\vspace{-.3em}
\section{Conclusion and Future Work}
\vspace{-.2em}
\label{sec:Conclusion}
{
\fullyjustified
When guiding \ltbqp query engines,
document linking structure and content policies
can vastly reduce the number of considered documents and triples,
yet for different reasons.
\linebreak
The usage of structure has \emph{performance} as a~goal,
so changing results is likely undesired;
\linebreak
content policies aim for \emph{result alterations},
thereby possibly eliminating network requests.
\linebreak
Since the effects occur
because of differences across contexts
rather than modifications to the query,
guided \ltbqp leads to
a~\emph{context-dependent notion of result completeness}.
\linebreak
Whereas the same query can thus have different results
in different user contexts,
the results are considered complete
in relation to the user-bound notion of \emph{relevant} results.

One of our main goals is bringing \ltbqp techniques
to competitive levels of performance
for cases where it would be challenging
to gather the data in a~query interface.
Concretely, we target decentralized networks of personal data,
where each user stores their own data
in their personal data vault,
guarded by access control~\cite{verborgh_timbl_chapter_2020}.
Given that orthogonal data interface features
such as access control and versioning
are much more straightforward to provide
on simple document-based interfaces
rather than more complex database-driven \apis,
there exists a~strong case for document-based query techniques.

This article only provides an initial sketch of the problem space.
Future work is required on the description
of linking structures and content policies.
We need a~detailed study of guided \ltbqp semantics
and its relation to other techniques and languages,
as~well as theoretical and empirical evaluations of its performance
in different contexts.
We plan to use the Comunica query engine platform~\cite{taelman_iswc_2018}
as a~basis for experimentation.

}

\vspace{-.5em}
\renewcommand\small{\fontsize{7.5}{9.33}\selectfont}
\bibliographystyle{splncs04}
\bibliography{references}
\end{document}